\documentclass[a4paper,11pt]{article}
\usepackage{jheppub} 
\title{\boldmath Dark Matter Admixed White Dwarfs: A Single-Fluid Approach}
\author[a]{Rajasmita Sahoo\,\orcidlink{0000-0002-9265-5025}}
\author[a]{, Somnath Mukhopadhyay\,\orcidlink{0000-0003-2666-3200}}
\affiliation[a]{National Institute of Technology, Tiruchirappalli, Tamil Nadu - 620015, India.}
\author[b]{and Mrutunjaya Bhuyan\,\orcidlink{0000-0002-8677-4220} \note{Corresponding author}}
\affiliation[b]{Institute of Physics, Sachivalaya Marg, Bhubaneswar, Odisha-751005, India. }
\emailAdd{rsphysics58@gmail.com}
\emailAdd{somnath@nitt.edu}
\emailAdd{mrutunjaya.b@iopb.res.in}

\abstract{In this study, we investigate the influence of an admixed fermionic dark matter (DM) component on the equilibrium structure of white dwarfs (WDs), with particular emphasis on the effects of varying the DM particle mass ($m_{\rm DM}$) and DM fraction ($f_{\rm DM}$). Notably, we employ a single-fluid approximation for the first time in this context, wherein the baryonic and DM contributions to the total energy density and pressure are treated within a unified framework, assuming non-interacting fermionic DM in hydrostatic equilibrium with baryons. We examine how variations in $m_{\mathrm{DM}}$ and $f_\mathrm{DM}$ modify the equation of state (EoS), the mass-radius relationship, and the internal mass and pressure distributions of WDs. Our results show that the presence of DM softens the EoS, with lighter DM particles providing stronger pressure support and leading to more extended stellar structures. Increasing the DM mass fraction leads to a more compact configuration, reducing both the radius and maximum mass of the WD. We further demonstrate that heavier DM particles enhance stellar compactness and can eventually drive the star toward gravitational instability. Moreover, the analysis of mass-radius relationships reveals that while small fractions of DM are consistent with observed WD masses, the radii predicted by our models are smaller than observations, suggesting additional influences such as rotation or magnetic fields. Our stability analysis confirms that the inclusion of dark matter does not lead to instability within the expected parameter space, indicating that white dwarfs admixed with dark matter can remain dynamically stable under certain conditions. These findings show that even a small admixture of DM can modify the structural properties and stability limits of WDs, providing a potential indirect astrophysical probe of DM particle properties.}

\begin{document}
\maketitle
\flushbottom
\section{Introduction}
\label{sec:intro}
White dwarfs (WDs) represent the final evolutionary state of low- and intermediate-mass stars (initial masses $\lesssim 8~M_{\odot}$), where nuclear fusion has ceased and gravitational collapse is prevented by electron degeneracy pressure. These compact remnants typically have masses around $0.6~M_{\odot}$ and radii comparable to that of Earth, with their structure described effectively by Chandrasekhar’s seminal work, which establishes an upper mass limit of approximately $1.44~M_{\odot}$ for non-rotating, non-magnetized WDs composed of carbon-oxygen nuclei and a degenerate electron gas~\cite{chandra35,shapiro83}. However, in recent years, observations of anomalous white dwarfs—exhibiting unusual mass-radius relationships, unexpected luminosities, or even masses exceeding the Chandrasekhar limit—have prompted a re-examination of the standard theoretical picture~\cite{andrew2006type, scalzo2010nearby, das13}. These discrepancies motivate the exploration of beyond-standard-model physics, with dark matter (DM) being a prominent candidate. The possibility that DM might influence stellar structure, either through direct capture over time or through primordial admixture, opens new avenues to constrain its properties using compact astrophysical objects.
The exact nature of DM remains one of the most pressing open questions in astrophysics and cosmology. Nonetheless, its existence is strongly supported by a wide range of observations, including galactic rotation curves~\cite{rubin70}, gravitational lensing~\cite{clowe06}, large-scale structure formation~\cite{springel05}, and cosmic microwave background (CMB) observations~\cite{planck18}. Numerous dark matter (DM) candidates have been proposed, including weakly interacting massive particles (WIMPs), axions, sterile neutrinos, and self-interacting asymmetric DM~\cite{bertone05,feng10}. In the astrophysical context, compact stars such as white dwarfs (WDs) provide unique laboratories for testing DM models. DM particles—particularly non-annihilating fermions—may become gravitationally bound to stars through capture or accretion processes~\cite{kouvaris10,goldman89}. Once captured, these particles can thermalise and form a DM core or halo within the star, thereby altering its structural equilibrium. In WDs, such DM admixture can modify the mass-radius relation, influence stability criteria, and change cooling timescales~\cite{leung13,NUNES2026100505,carvalho25,Graham18}. Even small DM cores with masses of order $10^{-6}$–$10^{-3}$~$M_\odot$ (depending on the DM particle mass) can produce detectable changes in WD structure~\cite{leung13}. More recently, two-fluid models have been employed to study the impact of DM on WD equilibrium~\cite{carvalho25,leung13,parmar25}. These studies reveal that even low-mass, asymmetric DM particles (with masses $\sim$10–100 GeV) can significantly lower the Chandrasekhar mass limit and influence electron-capture thresholds relevant for electron-capture supernovae (ECSNe). For instance, DM admixture can shift the critical mass for electron captures on $^{20}$Ne and $^{24}$Mg by $\Delta M \sim 0.01$–$0.05~M_\odot$, potentially triggering premature collapse~\cite{parmar25,takahashi13}.
In standard stellar evolution, progenitor stars in the mass range 8–10~$M_\odot$ evolve into super-asymptotic giant branch (SAGB) stars with degenerate oxygen–neon–magnesium (ONeMg) cores. When the core reaches a mass threshold of $\sim$1.36~$M_\odot$, electron captures can initiate a central O–Ne deflagration, resulting in either a core-collapse ECSN (producing a neutron star) or a failed explosion forming an ONeFe WD~\cite{jones19,wanajo10,podsiadlowski04}. The presence of DM can alter both the timing and nature of this transition, potentially contributing to the observed diversity in neutron star masses, spin distributions, and supernova energetics~\cite{muller14,janka12,bell20}. Thus, WDs serve not only as the end state of stellar evolution but also as sensitive probes of DM physics. The impact of DM admixture on stellar equilibrium, combined with observational constraints on WD masses and radii, provides a powerful tool for constraining DM properties, such as particle mass, interaction cross-section, and self-interaction strength. In this study, we investigate the equilibrium structure of white dwarfs (WDs) admixed with fermionic, non-annihilating dark matter (DM) within the framework of a single-fluid formalism. Notably, this work represents the first application of the single-fluid approach to model DM admixture in white dwarfs. We construct the equation of state (EoS) by incorporating the pressure contributions from both degenerate electrons and fermionic DM, and numerically solve the stellar structure equations. Our primary objective is to examine how the DM particle mass and fractional content influence the physical characteristics of WDs, and to explore the broader astrophysical implications for stellar collapse, supernova mechanisms, and compact object formation. The paper is organized as follows. In Sec.~\ref{sec:eoswd}, we review the equation of state (EoS) for ordinary degenerate matter in WDs. Sec.~\ref{sec:eosdm} introduces the EoS for fermionic dark matter. Sec.~\ref{sec:tov} presents the stellar structure equations and our numerical solution methods. In Sec.~\ref{sec:results}, we discuss the results, emphasizing the impact of DM admixture on WD properties. Sec.~\ref{sec:summary} concludes the paper with a summary and outlook.
\section{Equation of State of Ordinary Degenerate Matter in White Dwarfs} 
\label{sec:eoswd}
The equilibrium structure of a white dwarf (WD) is governed by the balance between the inward pull of gravity and the outward degeneracy pressure of the electron gas. Quantitatively, this balance is described by the equation of state (EoS), which relates the pressure to the energy density. At the high densities characteristic of WDs, the electrons are strongly degenerate and can be accurately modeled as a relativistic Fermi gas at zero temperature. To maintain charge neutrality, each proton must be accompanied by a electron, and together with neutrons these form nuclei. The baryonic component contributes the bulk of the mass density but exerts negligible pressure at zero temperature. In contrast, the electrons—being both degenerate and relativistic at sufficiently high densities—dominate the total pressure. \\
The electron number density is related to the Fermi momentum ($k_{F,e}$) as, 
\begin{equation}
    n_e = \frac{k_{F, e}^3}{3\pi^2\hbar^3}.
    \label{eq1}
\end{equation}
Since each electron is accompanied by a proton, the mass density can be written as,
\begin{equation}
    \rho = \frac{A}{Z}m_N n_e,
    \label{eq2}
\end{equation}
where, $m_{N}$ is the nucleon mass and ($A/Z$) denotes the average mass-to-charge ratio of the stellar composition (typically $A/Z=2$ for carbon-oxygen white dwarfs). \\
The dimensionless Fermi momentum is defined as,
\begin{equation}
    x_e=\frac{k_{F,e}}{m_e c}=\frac{\hbar}{m_e c}\left(3\pi^2\frac{Z}{A}\frac{\rho}{m_N}\right)^{1/3}.
    \label{eq3}
\end{equation}
With this parametrization, the energy density and pressure of the degenerate electron gas are given by~\cite{sagert06},
\begin{eqnarray}
\epsilon_{e}(k_{F,e}) &=& \frac{8\pi}{(2\pi\hbar)^3}\int_{0}^{k_{F,e}} (k^2c^2+m_e^2c^4)^{1/2}  k^2 dk, \nonumber 
\\
\epsilon_{e}(k_{F,e}) &=&\left(\frac{m_e^4 c^5}{8\pi^2\hbar^3}\right)\left[(2x_{e}^3+x_{e})(1+x_{e}^2)^{1/2}-sinh^{-1}(x_{e})\right], 
\label{eq4}
\end{eqnarray}
\begin{eqnarray}
P_{e}(k_{F,e})&=&\frac{1}{3}\frac{8\pi}{(2\pi\hbar)^3}\int_{0}^{k_{F,e}} \frac{k^2c^2}{(k^2c^2+m_{e}^2c^4)^{1/2}} k^2dk, \nonumber \\
P_{e}(k_{F,e}) &=& \left(\frac{m_e^4 c^5}{24\pi^2\hbar^3}\right)\left[(2x_{e}^3-3x_{e})(1+x_{e}^2)^{1/2}+3sinh^{-1}(x_{e})\right], 
\label{eq5}
\end{eqnarray}
By combining the electron contribution with the baryonic rest-mass term, the complete equation of state (EoS) of white dwarf can be written as,
\begin{eqnarray}
    \epsilon_{WD} = \epsilon_{e}(k_{F,e}) + \frac{A}{Z} m_{N}n_{e}c^2, \; P_{WD} = P_{e}(k_{F,e}).
\label{eq6}
\end{eqnarray}
This expression highlights that the baryonic component contributes predominantly to the rest-mass part of the total energy density, whereas the degenerate electrons provide the entire pressure support against gravitational collapse.
\section{Equation of State of Fermionic Dark Matter} 
\label{sec:eosdm} 
In the single-fluid formalism, we consider a degenerate gas of femionic dark matter (DM) particles that interacts with ordinary matter inside the white dwarf through the Higgs-portal mechanism. The DM and ordinary matter (baryonic matter) components are assumed to be well mixed and co--moving, with no relative flow. The DM particles are assumed to be non--annihilating and non--self-interacting, contributing additional pressure and energy density to the total stellar medium. The interaction between DM and baryonic matter is mediated by the Standard Model Higgs field, introducing a coupling that modifies the effective mass of the DM particle inside the dense stellar environment. This coupling enables a weak but non-negligible interaction between the two components, thereby altering the effective equation of state (EoS). The interacting Lagrangian density for this system, similar to that of nuclear matter in neutron star interacts with dark matter through the Standard Model Higgs-portal is given as~\cite{pano17,routaray23}:
\begin{eqnarray}
\mathcal{L}&=&\mathcal{L}_{\chi}+\mathcal{L}_{h}+\mathcal{L}_{baryon}+\mathcal{L}_{int}, \nonumber \\
\mathcal{L}_{\chi}&=&\bar{\chi}(i\gamma^{\mu}\partial_{\mu}-m_{\chi})\chi, \nonumber \\
\mathcal{L}_{h}&=&\frac{1}{2}\partial_{\mu}h\partial^{\mu}h-\frac{1}{2}m_{h}^2 h^2,  \nonumber \\
\mathcal{L}_{baryon} &=& \bar{\psi}_{e}(i\gamma^{\mu}\partial_{\mu}-m_{e})\psi_{e}+\bar{\psi}_{N}(i\gamma^{\mu}\partial_{\mu}-m_{N})\psi_{N},\nonumber \\
\mathcal{L}_{int} &=& y\bar{\chi}\chi h + g_{hN} \bar{\psi}_{N}\psi_{N}h+g_{he}\bar{\psi_{e}}\psi_{e}h. 
\label{eq7}
\end{eqnarray}
Here, $\chi$ represents the fermionic DM field with mass $m_{\chi}$, and $h$ denotes the Higgs field with mass $m_{h} = 125\,\mathrm{GeV}$. The symbols $\psi_{e}$ and $\psi_{N}$ correspond to the electron and nucleon fields, respectively. The parameter $y$ denotes the DM–Higgs Yukawa coupling, while the Higgs–nucleon coupling is expressed as $g_{hN} = f m_{N} / v$, where $f$ is the nucleon form factor and $v$ is the Higgs vacuum expectation value. Although the Higgs–electron coupling $g_{he}$ is present, it is negligibly small compared to $g_{hN}$. In a static and dense medium, the Higgs field is commonly replaced by its mean-field expectation value, $h \rightarrow h_{0}$, leading to an effective dark matter mass given by
\begin{equation}
    m_{\chi}^{eff}=m_{\chi}-y h_{0}. 
    \label{eq8}
\end{equation}
From this point onwards, we redefine the notation for fermionic dark matter particle from $\chi$ to DM. This notation is maintained throughout the equations of state and stellar structure analysis.
The Fermi momentum of the dark matter component is given as 
\begin{equation}
    k_{F,{DM}}=(3\pi^2 n_{DM})^{1/3},
    \label{eq9}
\end{equation}
where $n_{DM}$ is the DM number density. In this study, $k_{F,DM}$ is varies from $0$ to $0.05\; \mathrm{GeV}$ to analyze the impact of dark matter on the global properties of white dwarf, while the dark matter particle mass $(m_\mathrm{DM})$ is considered in the range $0.1 - 10\; \mathrm{GeV}$. \\
The energy density and pressure of the fermionic DM component are obtained as follows:
\begin{equation}
\epsilon_{DM} = \frac{2}{(2\pi)^3}\int_{0}^{k_{F,DM}} \sqrt{k^2+(m_{DM}^{eff})^2} d^3k\; + \epsilon_{h}(h_{0}),
\label{eq10}
\end{equation}
\begin{equation}
    P_{DM}=\frac{2}{3(2\pi)^3}\int_{0}^{k_{F,DM}}\frac{k^2}{\sqrt{k^2+(m_{DM}^{eff})^2}} d^3k - \epsilon_{h}(h_{0}).
    \label{11}
\end{equation}
where $\epsilon_{h} (h_{0}) = \frac{1}{2} m_{h}^2h_{0}^2$ is the Higgs mean-field energy. \\
The white dwarf matter is modeled using the conventional degenerate Fermi gas formalism discussed in Sec.~\ref{sec:eoswd}. Within the single-fluid approach, both the white dwarf (degenerate electrons + nucleons) and the DM component are treated as a single effective fluid characterized by total energy density and total pressure given by
\begin{eqnarray}
    \epsilon_{T}=\epsilon_{WD}+\epsilon_{DM}, \; P_{T} = P_{WD} + P_{DM}.
    \label{eq12}
\end{eqnarray} 
In this approach, the dark matter fraction ($f_\mathrm{DM}$) represents the fractional contribution of DM to the total number density of the system (white dwarf-dark matter admixture). It is defined as
\begin{equation}
    f_{DM} = \frac{n_{DM}}{(n_{WD}+n_{DM})},
    \label{eq13}
\end{equation}
where $n_\mathrm{DM}$ and $n_\mathrm{WD}$ are the respective number densities. In the present formalism, the dark matter fraction is assumed to lie within the range $f_\mathrm{DM}= (0.01 - 0.1)$. The corresponding total number density of the admixed configuration is then,
\begin{equation}
    n_{T} = n_{WD} + n_{DM}.
    \label{eq14}
\end{equation}
This formalism ensures hydrostatic equilibrium under a common gravitational potential, making it suitable for describing systems in which dark matter and baryonic matter are well mixed. The resulting effective total EoS $P_{T}(\epsilon_{T})$ is incorporated into the Tolman-Oppenheimer-Volkoff (TOV) equations to obtain the stellar structure of dark matter-admixed white dwarfs. In the following sections, we compute the EoS for various choices of $m_\mathrm{DM}$ and $f_\mathrm{DM}$ and systematically examine their impact on the global stellar configurations. The inclusion of the Higgs-mediated DM component tends to soften the total EoS, leading to smaller radii and reduced maximum masses for higher $f_\mathrm{DM}$ or larger $m_\mathrm{DM}$.
\section{Stellar Structure Equations for Dark Matter-Admixed White Dwarfs  }
\label{sec:tov}
The equilibrium configuration of a compact star, such as a white dwarf (WD) admixed with dark matter (DM), is governed by the Tolman–Oppenheimer–Volkoff (TOV) equations, which describe hydrostatic equilibrium within the framework of general relativity. In the single-fluid formalism, both the ordinary degenerate matter of the WD and the DM component are treated as a unified effective fluid characterized by a total pressure $P_{T}(r)$ and total energy density $\epsilon_{T}(r)$. The corresponding stellar structure equations are given by~\cite{tolman39,oppen39},
\begin{equation}
    \frac{dP_{T}(r)}{dr}=-\frac{Gm(r)\epsilon_{T}(r)}{r^2}\left(1+\frac{P_{T}(r)}{\epsilon_{T}(r)c^2}\right) 
    \left(1+\frac{4\pi r^3P_{T}(r)}{m(r)c^2}\right)\left(1-\frac{2G m(r)}{rc^2}\right)^{-1},
    \label{eq15}
\end{equation}
\begin{equation}
    \frac{dm(r)}{dr}=4\pi r^2 \epsilon_{T}(r).
    \label{eq16}
\end{equation}
where $G$ is the gravitational constant, $c$ is the speed of light and $m (r)$ is the enclosed mass at radial coordinate $r$. 
In the single-fluid approach, the total EoS incorporates contributions from both the ordinary WD matter and the degenerate fermionic DM component. The energy density $\epsilon_{\mathrm{WD}}$ and pressure $P_{\mathrm{WD}}$ of the WD matter are obtained from the EoS described in Sec.~\ref{sec:eoswd}, while the DM energy density $\epsilon_{\mathrm{DM}}$ and pressure $P_{\mathrm{DM}}$ follow from the fermionic DM EoS presented in Sec.~\ref{sec:eosdm}. The coupled differential equations (~\ref{eq15})–(~\ref{eq16}), together with the composite EoS given in Eq.~(\ref{eq12}), fully describe the structure of a DM-admixed white dwarf within the single-fluid framework. For a specified central total energy density $\epsilon_{T0} = \epsilon_{T} (r=0)$ and central total pressure $P_{T0} = P_{T} (r=0)$, obtained from the effective EoS [Eq.~(\ref{eq12})], the TOV equations [Eqs.~(\ref{eq15}) and ~(\ref{eq16})] are integrated outward from the center. The integration continues until the pressure drops to zero, $P_{T}(r=R) = 0$, which defines the stellar radius $R$. The total gravitational mass is then given by $M = m(r=R)$, thereby establishing the mass–radius relation for the equilibrium configurations of DM–admixed white dwarf.
\begin{center}
\begin{table}[t]
\centering
\renewcommand{\arraystretch}{1.3} 
\setlength{\tabcolsep}{10pt}      
\begin{tabular}{|ccc||ccc|}
\hline \hline
\multicolumn{3}{|c||}{{\bf $m_{\mathrm{DM}}$ = 1 GeV}} & 
\multicolumn{3}{|c|}{{\bf $f_{\mathrm{DM}}$ = 0.05}} \\
\hline
$f_{\mathrm{DM}}$ & $M/M_\odot$ & $R$ (km) & $m_{\mathrm{DM}}$ & $M/M_\odot$ & $R$ (km) \\ 
\hline
0.00&1.40&1020 & 0.1 & 1.43 & 935\\
0.01 & 1.39 & 1019 & 1.0 & 1.36 & 1018 \\ 
0.05 & 1.36 & 1018 & 2.0 & 1.29 & 951  \\ 
0.07 & 1.34 & 976  & 3.0 & 1.22 & 927   \\ 
0.08 & 1.33 & 975  & 4.0 & 1.16 & 903\\  
0.10  & 1.32 & 974  & 5.0 & 1.11 & 881   \\ 
\cline{1-3} 
\multicolumn{3}{|c||}{{\bf $m_{\mathrm{DM}}$ = 10 GeV}} & 6.0 & 1.05 & 860\\ 
\cline{1-3}
0.00&1.40&1020& 7.0 & 1.01 & 840 \\
0.01 & 1.27 & 973 & 8.0 & 0.96 & 792 \\ 
0.05 & 0.88 & 757 & 9.0 & 0.92 & 774  \\
0.07 & 0.74 & 678 & 10.0 & 0.88 &757  \\  
0.08 & 0.68 & 653 & & &\\ 
0.10 & 0.59 & 607 &  & &\\   
\hline \hline
\end{tabular}
\caption{Variation of the maximum mass ($M$) and corresponding stellar radius ($R$) of dark matter (DM)–admixed white dwarfs. Left part: Dependence of $M$ and $R$ on the dark matter fraction ($f_\mathrm{DM}$), computed using a single-fluid approach for fixed dark matter particle masses $m_\mathrm{{DM}} = 1\;\mathrm{GeV}$ and $10\;\mathrm{GeV}$. Right part: Variation of $M$ and $R$ with different dark matter particle masses ($m_\mathrm{DM}$) for a fixed dark matter fraction $f_\mathrm{{DM}} = 0.05$. The calculated masses are consistent with observed white dwarfs ZTF J1901+1458 ($1.33 \leq M/M_\odot \leq 1.37$)~\cite{Cai21}, Sirius B ($M/M_\odot = 1.018 \pm 0.011$)~\cite{Bond17}, Stein 2051 B ($M/M_\odot = 0.675 \pm 0.051$) \cite{Sahu17}, 40 Eridani B ($M/M_\odot = 0.573 \pm 0.018$)~\cite{bond2017}, and GK Vir ($M/M_\odot = 0.56 \pm 0.01$)~\cite{parson17}.} 
\label{tab1}
\end{table}
\end{center}
\section{Results and Discussion} 
\label{sec:results}
In this study, we investigate the effect of an admixed fermionic dark matter (DM) component on the equilibrium structure of white dwarfs (WDs), with particular emphasis on the roles of the DM particle mass ($m_{\rm DM}$) and DM mass fraction ($f_{\rm DM}$). A single-fluid approximation is adopted, in which the total energy density and pressure include contributions from both baryonic matter and DM. To the best of our knowledge, this work represents the first application of the single-fluid approach in model dark matter–admixed white dwarfs. This approximation is valid under the assumption that the fermionic dark matter is non--self-interacting, non-annihilating, and remains in hydrostatic equilibrium with the baryonic matter. Within this formulation, variations in $m_{\rm DM}$ and $f_{\rm DM}$ lead to significant modifications in the total equation of state (EoS), thereby altering the internal pressure and mass distributions. The possible maximum mass and dark matter (DM) fraction that can be sustained within a white dwarf (WD) are strongly constrained by both theoretical stability limits and astrophysical observations. Earlier studies have shown that the addition of an admixed DM component can substantially alter the stellar structure by softening the total equation of state (EoS), thereby reducing the maximum stable mass of the WD~\cite{leung13,carvalho25,parmar25}. For light fermionic DM particles ($m_{\rm DM} \lesssim 1$–$10\;\mathrm{GeV}$), the DM pressure contribution can partially counteract gravitational compression, allowing a small DM mass fraction ($f_{\rm DM} \sim 0.01$–$0.05$) to coexist with the baryonic component in hydrostatic equilibrium~\cite{leung13,carvalho25,parmar25}. However, for heavier DM particles ($m_{\rm DM} \gtrsim 10\;\mathrm{GeV}$), the gravitational dominance of the DM component leads to a more compact configuration and a pronounced reduction in the Chandrasekhar mass limit, potentially down to $M_{\rm max} \lesssim 1.0\,M_\odot$ for $f_{\rm DM} \approx 0.05$~\cite{parmar25}. Observationally, the existence of ultra-massive white dwarfs such as ZTF J1901+1458 ($M/M_\odot \approx 1.35$)~\cite{Cai21} provides stringent constrains on the amount of DM  that can be admixed. These measurements suggest that any significant DM contribution must be limited to only a few percent of the total stellar mass to remain consistent with current WD mass-radius observations. Hence, both theoretical modeling and astrophysical evidence indicate that dark matter fractions exceeding $f_{\rm DM} \sim 0.1$ are unlikely to yield stable WD configurations, establishing an upper bound on both the DM particle mass and the total DM contribution that can be supported within a white dwarf framework. 
\begin{figure*}[t]
\centering
\includegraphics[width=0.95\textwidth]{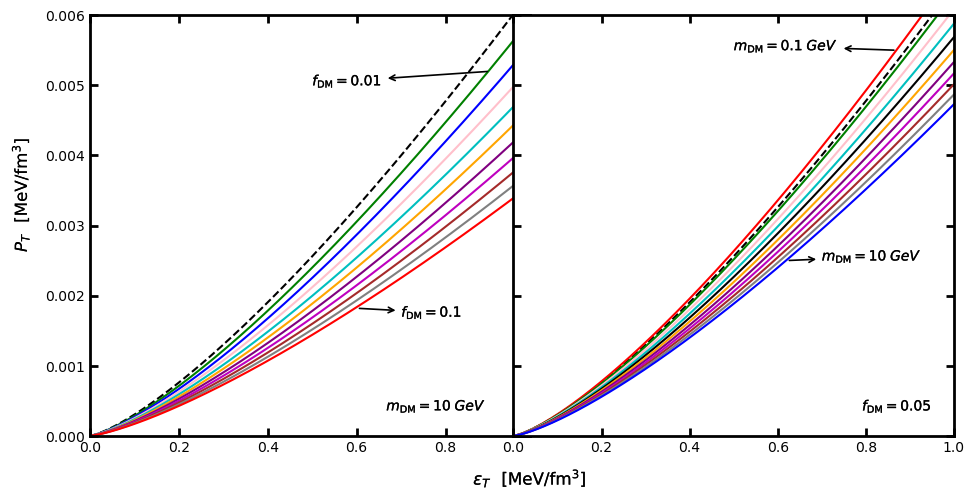}
\caption{Variation of the equation of state (EoS) for dark matter-admixed white dwarfs. Left panel: Total pressure $P_{T}$ as a function of total energy density $\epsilon_{T}$ for a fixed dark matter particle mass $m_\mathrm{DM}=10\,\mathrm{GeV}$. The arrows mark the curves corresponding to $f_\mathrm{DM}=0.01$ and $f_\mathrm{DM}=0.1$, while the intermediate curves represent the other dark matter fractions $f_\mathrm{DM}= [0.02,0.03,0.04,0.05,0.06,0.07,0.08,0.09]$. Right panel: The variation of the EoS for a fixed dark matter fraction $f_\mathrm{DM}=0.05$. The arrows indicate the curves corresponding to $m_\mathrm{DM}=0.1\,\mathrm{GeV}$ and $m_\mathrm{DM}=10\,\mathrm{GeV}$, with the intermediate curves showing the results for $m_\mathrm{DM}=[1,2,3,4,5,6,7,8,9]\,\mathrm{GeV}$. In both panels, the dashed black line represents the standard Chandrasekhar white dwarf without dark matter.} 
\label{fig1}
\end{figure*}
\begin{figure*}[t]
    \centering
    \includegraphics[width=0.95\textwidth]{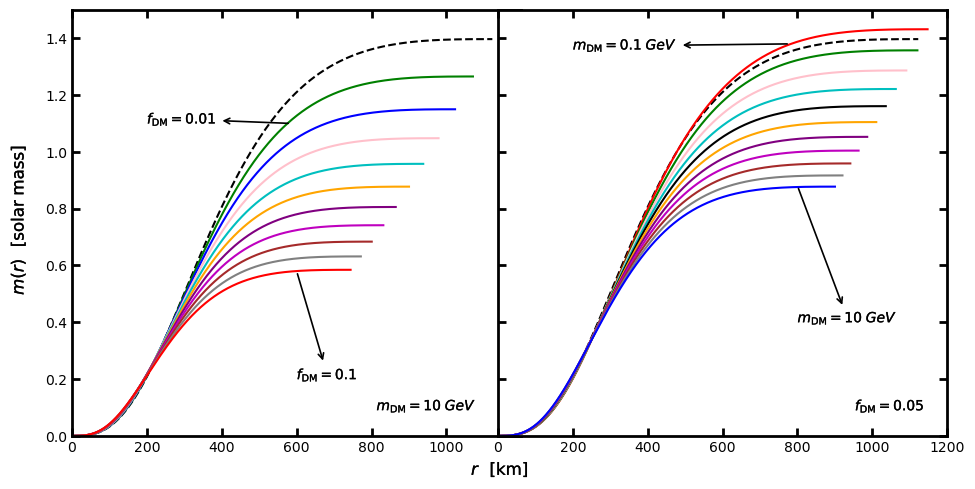}
    \caption{Radial mass profile $m(r)$ of dark matter-admixed white dwarfs. Left panel: The mass distribution for a fixed dark matter particle mass $m_\mathrm{DM}$=10\;GeV. The arrows indicate the curves corresponding to $f_\mathrm{{DM}}=0.01$ and $f_\mathrm{{DM}}=0.1$, while the intermediate curves represent other dark matter mass fractions $f_\mathrm{{DM}}=[0.02,0.03,0.04,0.05,0.06,0.07,0.08,0.09]$. Right panel: The radial mass profile for a fixed dark matter mass fraction $f_\mathrm{{DM}}=0.05$. The arrows mark the curves corresponding to $m_\mathrm{{DM}}=0.1\;\mathrm{GeV}$ and $m_\mathrm{{DM}}=10\;\mathrm{GeV}$, with the intermediate curves corresponding to other dark matter particle masses $m_\mathrm{{DM}}=\mathrm{[1,2,3,4,5,6,7,8,9]\;GeV}$. In both panels, the dashed black line represents the standard Chandrasekhar white dwarf without dark matter. }
    \label{fig2}
\end{figure*} 
\begin{figure*}[t]
    \centering
    \includegraphics[width=0.95\textwidth]{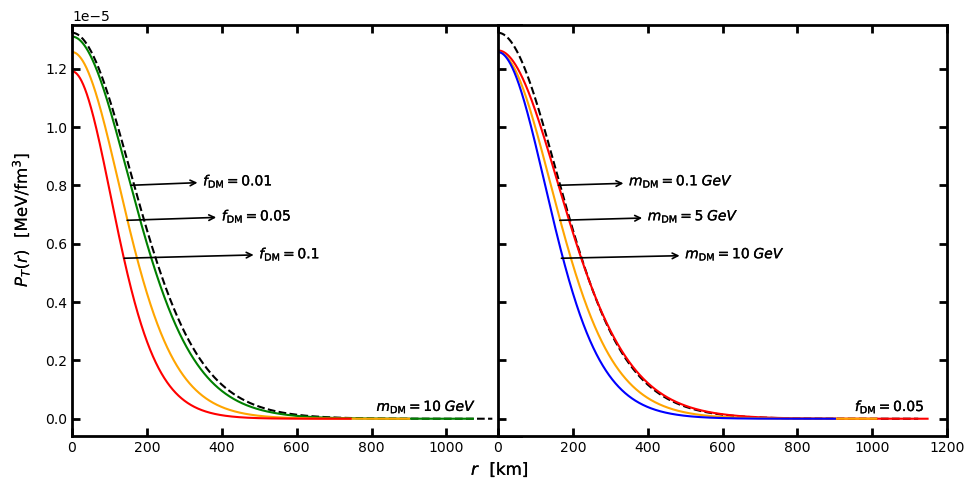}
    \caption{Radial pressure profile $P_{T}(r)$ of dark matter-admixed white dwarfs. Left panel: The pressure distribution for a fixed dark matter particle mass $m_\mathrm{{DM}}=\mathrm{10\;GeV}$. The arrows indicate the curves corresponding to $f_\mathrm{{DM}}=0.01$, $f_\mathrm{{DM}}=0.05$, and $f_\mathrm{{DM}}=0.1$. Right panel: The pressure profile for a fixed dark matter mass fraction $f_\mathrm{{DM}}=0.05$. The arrows mark the curves corresponding to $m_\mathrm{{DM}}=\mathrm{0.1\;GeV}$, $m_\mathrm{{DM}}=\mathrm{5\;GeV}$, and $m_\mathrm{{DM}}=\mathrm{10\;GeV}$. In both panels, the dashed black line denotes the standard Chandrasekhar white dwarf without dark matter.}
    \label{fig3}
\end{figure*} 
\begin{figure*}[t]
\centering
\includegraphics[width=0.95\textwidth]{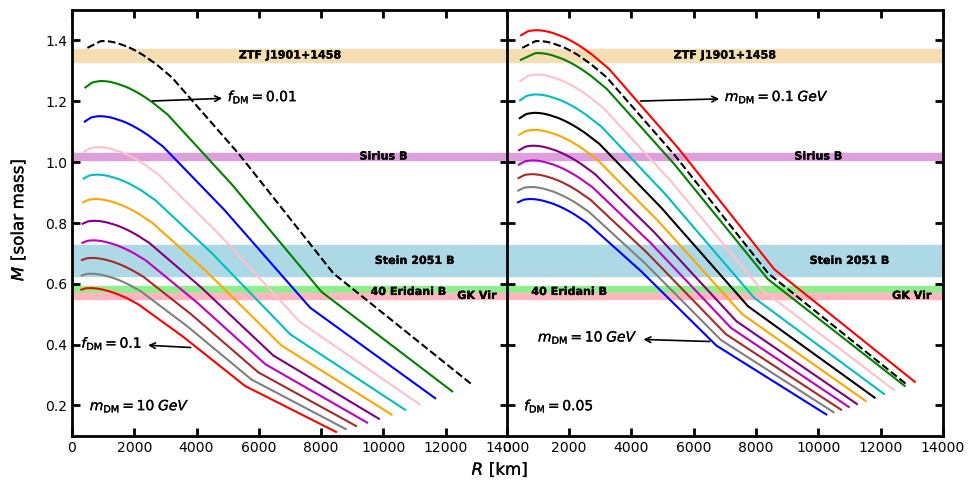}
\caption{Mass-radius relationship of dark matter-admixed white dwarfs. Left panel: The mass-radius curves for a fixed dark matter particle mass $m_\mathrm{{DM}}=\mathrm{10\;GeV}$, where the arrows indicate the curves corresponding to $f_\mathrm{DM}=0.01$ and $f_\mathrm{DM}=0.1$; the intermediate curves correspond to other dark matter mass fractions $f_\mathrm{DM}=[0.02,0.03,0.04,0.05,0.06,0.07,0.08,0.09]$. Right panel: The mass-radius relationship for a fixed dark matter mass fraction $f_\mathrm{DM}=0.05$, with arrows marking the curves for $m_\mathrm{{DM}}=\mathrm{0.1\;GeV}$ and $m_\mathrm{{DM}}=\mathrm{10\;GeV}$; the intermediate curves correspond to other dark matter particle masses $m_\mathrm{{DM}}=\mathrm{[1,2,3,4,5,6,7,8,9]\;GeV}$. In both plots, the dashed black line represents the standard Chandrasekhar white dwarf without dark matter, and no legend is displayed. Shaded horizontal bands indicate recent observational mass constraints from $\mathrm{ZTF\;1901+1458}$ (wheat), $\mathrm{Sirius\;B}$ (plum), $\mathrm{Stein\; 2051\;B}$ (light blue), $\mathrm{40\;Eridani\;B}$ (light green), and $\mathrm{GK\;Vir}$ (light pink), representing the range of astrophysically viable stellar configurations.}
\label{fig4}
\end{figure*} 

\subsection{Equation of State of Dark Matter-Admixed White Dwarfs: Dependence on $m_{\rm DM}$ and $f_{\rm DM}$} 
We first examine the effect of the DM particle mass by fixing the DM mass fraction at $f_{\rm DM} = 0.05$ and varying the DM particle mass in the range $m_{\rm DM} = 0.1$–$10$ GeV. The choice of $f_{\rm DM} = 0.05$ is motivated by earlier studies~\cite{carvalho25,parmar25}, which suggest that even a small admixture of dark matter within a compact star can produce measurable structural deviations without destabilizing the configuration. This specific fraction ensures that the DM component remains dynamically subdominant, thus preserving the overall baryonic character of the star, while allowing a physically meaningful investigation of how DM particle mass affects the macroscopic properties such as mass–radius relation and stability limits. Within this framework, the total EoS is obtained by summing the contributions from the degenerate electron gas and the non-interacting fermionic DM component. As shown in the right panel of Fig.~\ref{fig1} and in Table~\ref{tab1}, the presence of DM softens the EoS compared to a standard WD. This softening is less pronounced for lighter DM particles, which exhibit higher degeneracy pressure and thus greater pressure support at a given energy density. In contrast, heavier DM particles contribute comparatively lower pressure, leading to a more pronounced reduction in the total pressure. This behavior reflects the inverse relationship between degeneracy pressure and particle mass in a relativistic fermion gas. Next, by fixing the dark matter (DM) particle mass at $m_{\rm DM}$ = 10 GeV, we explore the influence of varying the DM mass fraction in the range $f_{\rm DM}$ = 0.01-0.1. The resulting EoSs, presented in the left panel of Fig.~\ref{fig1} and summarized in Table~\ref{tab1}, exhibit a progressive softening as $f_{\rm DM}$ increases. This arises because the DM component contributes less pressure per unit energy density compared to the degenerate electron gas that dominates the baryonic matter. Consequently, increasing DM fraction $f_\mathrm{DM}$, lower the total pressure at a given energy density, leading to an overall softer EoS. Physically, a larger DM admixture reduces the effective pressure support against gravitational collapse, leading to a more compact stellar structure. This manifests as a decrease in both the stellar radius and the maximum stable mass of the white dwarf. At sufficiently large $f_{\rm DM}$, the star becomes increasingly gravitationally bound, consistent with earlier theoretical predictions that DM enrichment enhances the central density and suppresses the Chandrasekhar mass limit~\cite{leung13,carvalho25,parmar25}. These findings further indicate that the structural response of a white dwarf to DM admixture is highly sensitive to the fractional DM content, providing an important probe for constraining the presence and properties of dark matter within compact stellar systems.  
\subsection{Radial Mass and Pressure Profiles for Varying $m_{\rm DM}$ and $f_{\rm DM}$}
The left panel of Fig. \ref{fig2} presents the radial mass distribution $m(r)$ for dark matter-admixed white dwarfs (DM-WDs) with dark matter fractions in the range $f_{\rm DM} = 0.01\text{–}0.1$, at a fixed DM particle mass of $m_{\rm DM} = 10~\mathrm{GeV}$. As $f_{\rm DM}$ increases, the stellar mass profile becomes increasingly concentrated toward the center, indicating a more compact core. Quantitatively, the mass enclosed within the inner $0.5 R$ of the star decreases by approximately $\sim 54\%$ as $f_{\rm DM}$ increases from 0.01 to 0.1, depending on the specific equation of state (EoS). This enhanced central concentration arises from the EoS softening induced by the DM component, which reduces the pressure support in the outer baryonic layers and strengthens the overall gravitational confinement. Similar behavior has been reported in earlier studies~\cite{leung13,carvalho25,parmar25}, where increasing DM content was shown to steepen the density gradient, shifts mass toward the stellar core, resulting in enhanced central density and reduces total stellar radius. \\
In the right panel of Fig.~\ref{fig2}, we fix the DM mass fraction at $f_{\rm DM} = 0.05$ and vary the DM particle mass over the range $m_{\rm DM} = 0.1\text{–}10~\mathrm{GeV}$. The resulting mass profiles demonstrate that lighter DM particles--due to their higher degeneracy pressure ($P_\mathrm{DM} \propto n_\mathrm{DM}^{5/3}/m_{\rm DM}$) in the non-relativistic limit--produce more extended stellar configurations. Conversely, as $m_{\rm DM}$ increases, the degeneracy pressure contribution diminishes, causing the stellar matter to contract further toward the center. For $m_{\rm DM} \gtrsim 5~\mathrm{GeV}$, the mass distribution becomes significantly peaked towards the center, leading to higher central densities and smaller radii, consistent with the trends predicted by Refs.~\cite{carvalho25,parmar25}. This quantitative relationship illustrates the inverse dependence of pressure support on particle mass, a fundamental property of fermionic systems that directly governs the structural compactness of DM-WDs. \\
To further clarify the structural response, we examine the radial total pressure profiles $P_T(r)$ shown in Fig.~\ref{fig3}. In the left panel, for $m_{\rm DM} = 10~\mathrm{GeV}$, the central pressure decreases systematically with increasing $f_{\rm DM} = 0.01, 0.05, 0.1$. The pressure gradients steepen as the DM content rises, indicating enhanced gravitational binding and stronger central condensation. In the right panel of Fig.~\ref{fig3}, where $f_{\rm DM} = 0.05$ is fixed and $m_{\rm DM}$ varies from 0.1 to 10 GeV, the central pressure exhibits a clear declining trend with increasing $m_{\rm DM}$. The reduction in degeneracy pressure for heavier DM species leads to steeper pressure gradients, thereby reinforcing the overall compactness of the configuration. These results are consistent with prior analyses that correlate the degree of DM-induced softening with both particle mass and fractional abundance~\cite{carvalho25,parmar25,Bartlett2022}. Overall, the trends presented in Fig.~\ref{fig2} and Fig.~\ref{fig3} demonstrate that heavier DM particles and higher DM fractions synergistically enhance the gravitational potential well, yielding denser and more compact white dwarfs compared to their purely baryonic counterparts. 
\subsection{Mass--Radius Relation for Varying $m_\mathrm{DM}$ and $f_\mathrm{DM}$}
The left panel of Fig.~\ref{fig4} (and Table~\ref{tab1}) illustrates the impact of varying dark matter (DM) mass fractions on the mass–radius ($M–R$) relation for white dwarfs (WDs). As the DM fraction increases from $f_{\rm DM} = 0$ to $f_{\rm DM} = 0.1$, the $M \sim R$ curves systematically shift toward smaller radii for a given stellar mass, indicating increasingly compact configurations. Quantitatively, for $f_{\rm DM} = 0.01$, the stellar radius decreases by approximately $\sim$ 5\%  relative to a standard WD of comparable mass, while the maximum stable mass reduces by nearly $\sim$ 9\% for the adopted equation of state (EoS). This trend arises because the DM component introduces additional gravitational binding and effectively softens the total EoS and diminishes the pressure support provided by the degenerate electron gas. At higher DM admixtures $f_{\rm DM} > 0.1$, both the stellar mass and radius continue to decline sharply. For sufficiently large $f_{\rm DM}$, the configuration becomes gravitationally unstable and collapses—consistent with earlier findings~\cite{leung13,carvalho25,parmar25}, which reported that excessive DM content drives the white dwarfs below the Chandrasekhar mass limit. These results establish an upper bound on the DM fraction that can be stably supported in WD–DM systems, typically around $f_{\rm DM} \lesssim 0.1$ for fermionic DM in the GeV mass range. The observed mass ranges for several well-known white dwarfs—such as ZTF J1901+1458 ($1.33 \leq M/M_\odot \leq 1.37$)~\cite{Cai21}, Sirius B ($M/M_\odot = 1.018 \pm 0.011$) \cite{Bond17}, Stein 2051 B ($M/M_\odot = 0.675 \pm 0.051$)~\cite{Sahu17}, 40 Eridani B ($M/M_\odot = 0.573 \pm 0.018$)~\cite{bond2017}, and GK Vir ($M/M_\odot = 0.56 \pm 0.01$)~\cite{parson17} are represented by shaded horizontal bands in Fig.~\ref{fig4} and listed in Table~\ref{tab1}. The theoretical predictions from our WD–DM models lie within these observational mass ranges, however, the corresponding radii are systematically smaller than observed values. Similar discrepancies have been reported in previous works~\cite{leung13,carvalho25}, suggesting that additional physical effects—such as stellar rotation, magnetic field support, or finite-temperature corrections—may be necessary to reconcile the model predictions with observed WD radii.
The right panel of Fig.~\ref{fig4} (and Table~\ref{tab1}) shows the variation in the $M \sim R$ relation with different DM particle masses $m_{\rm DM}$, at a fixed DM fraction of $f_{\rm DM} = 0.05$. For light DM particles ($m_{\rm DM} < 1~\mathrm{GeV}$), the mass–radius curve remains nearly indistinguishable from the standard Chandrasekhar relation, indicating that the pressure contribution from the DM component is negligible compared to the electron degeneracy pressure. However, for heavier DM particles ($m_{\rm DM} > 1~\mathrm{GeV}$), the configurations become progressively more compact, with smaller radii and reduced maximum masses. Quantitatively, as $m_{\rm DM}$ increases from 0.1 to 10 GeV, the maximum mass decreases by nearly $\sim 38\%$, and the corresponding radius contracts by up to $\sim 19\%$. For $m_{\rm DM} \gtrsim 10~\mathrm{GeV}$, the degeneracy pressure from the DM component becomes negligible, and the gravitational contribution of the DM component dominates, leading to dynamically unstable configurations that may undergo collapse. This trend is consistent with previous analyses~\cite{carvalho25, parmar25, Bartlett2022}, which showed that heavier DM particles enhance gravitational binding and accelerate the onset of instability in compact stars. The maximum masses and corresponding radii for each combination of $m_{\rm DM}$ and $f_{\rm DM}$ are summarized in Table~\ref{tab1}, quantitatively highlighting the sensitivity of WD structure to the properties of the DM component.
\subsection{Total Mass variation with Central Total Number Density and Energy Density }
\begin{figure*}[t]
\centering
\includegraphics[width=0.95\textwidth]{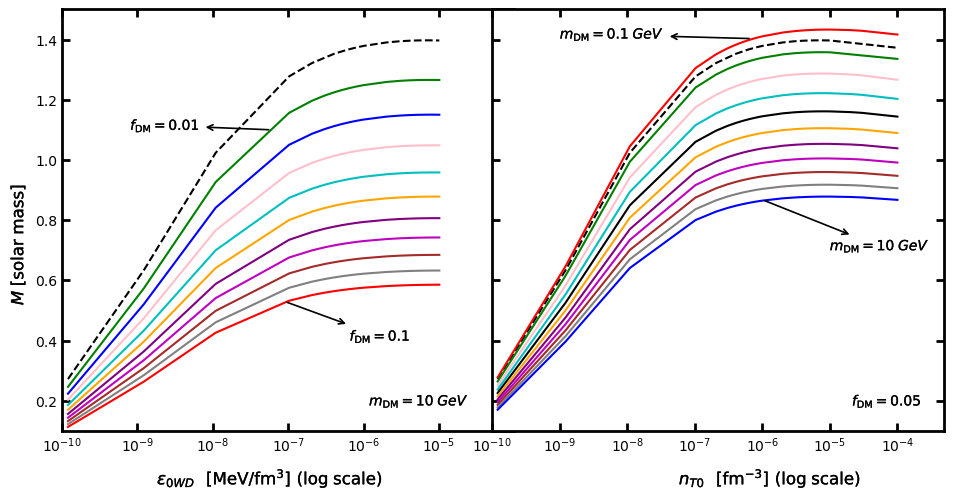}
\caption{Mass variation of dark matter-admixed white dwarfs. Left panel: The variation of stellar mass with central white dwarf energy density $\epsilon_{0WD}$ (in log scale) for a fixed dark matter particle mass $m_\mathrm{{DM}}=\mathrm{10\;GeV}$. The arrows mark the curves corresponding to $f_\mathrm{{DM}}=0.01$ and $f_\mathrm{{DM}}=0.1$, while the intermediate curves correspond to other dark matter mass fractions $f_\mathrm{{DM}}=\mathrm{[0.02,0.03,0.04,0.05,0.06,0.07,0.08,0.09]}$. Right panel: The variation of stellar mass with total central number density $n_{T0}$ (in log scale) for a fixed dark matter mass fraction $f_\mathrm{{DM}}=0.05$. The arrows indicate the curves corresponding to $m_\mathrm{{DM}}=\mathrm{0.1\;GeV}$ and $m_\mathrm{{DM}}=\mathrm{10\;GeV}$ with intermediate curves representing other dark matter particle masses $m_\mathrm{{DM}}=\mathrm{[1,2,3,4,5,6,7,8,9]\;GeV}$. In both panels, the dashed black line represents the standard white dwarf without dark matter, and no legend is displayed. }
\label{fig5}
\end{figure*}

\begin{figure}[t]
\vspace{0.0cm}
\eject\centerline{\epsfig{file=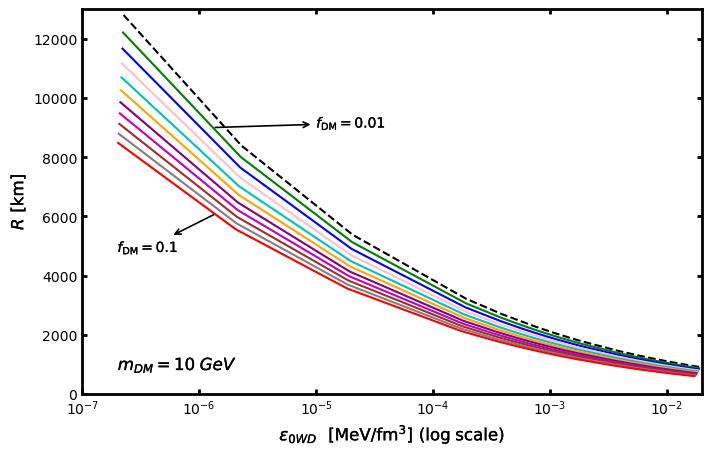,height=9cm,width=9cm}}
\caption{Radius as a function of the central energy density of white dwarf $\epsilon_{0WD}$ (in log scale) for dark matter-admixed white dwarfs, shown for a fixed dark matter particle mass $m_\mathrm{{DM}} =\mathrm{10\;GeV}$. Arrows indicate the curves corresponding to dark matter mass fractions $f_\mathrm{{DM}}=0.01$ and $f_\mathrm{{DM}}=0.1$, with intermediate curves representing other dark matter mass fractions $f_{DM}=[0.02,0.03,0.04,0.05,0.06,0.07,0.08,0.09]$. The dashed black line denotes the standard Chandrasekhar white dwarf without dark matter.}
\vspace{0.0cm}
\label{fig6}
\end{figure}

The right panel of Fig.~\ref{fig5} illustrates the variation of the total gravitational mass of dark matter-admixed white dwarfs (WD–DM systems) as a function of the central total number density ($n_{T0}$). As in the case of standard white dwarfs, the total mass initially increases with $n_{T0}$, reaches a maximum value, and subsequently decreases at higher densities—consistent with the canonical Chandrasekhar mass–density relation~\cite{Chandrasekhar1931}. Quantitatively, for the purely baryonic configuration ($f_{\rm DM} = 0$), the maximum mass occurs at approximately $n_{T0} \sim 7\times10^{-6}\,\mathrm{fm^{-3}}$, corresponding to a maximum mass of $M_{\rm max} \approx 1.4\,M_\odot$. When DM is included, the qualitative behavior of the mass-density curve remains similar, but the maximum mass shifts to lower values as either the DM particle mass ($m_{\rm DM}$) or the DM fraction ($f_{\rm DM}$) increases. For instance, at a fixed $n_{T0}$, increasing $m_{\rm DM}$ from 0.1 to 10~GeV results in a reduction of the total mass by approximately $\sim 38\%$, depending on the adopted EoS. This suppression occurs because heavier DM particles contribute negligibly to the total pressure (since $P_\mathrm{DM} \propto n_\mathrm{ DM}^{5/3}/m_{\rm DM}$), yet they add significant gravitational weight to the system, leading to stronger compression and smaller equilibrium masses. Similar mass suppression effects have been reported in earlier theoretical works~\cite{leung13,carvalho25,parmar25}, where it was demonstrated that DM enrichment steepens the mass–density curve and accelerates the onset of instability. Thus, the trend observed in Fig.~\ref{fig5} reflects the competition between the degeneracy pressure of the baryonic component and the additional gravitational potential contributed by the DM admixture.
Further extending the analysis for dependence of the total mass and radius of WD–DM systems on the central energy density of the baryonic component ($\epsilon_{0\rm WD}$) for varying dark matter mass fractions ($f_{\rm DM}$) and are shown in the left panel of Fig.~\ref{fig5} and Fig.~\ref{fig6}. The total mass initially increases with $\epsilon_{0\rm WD}$, attains a maximum, and then declines as $\epsilon_{0\rm WD}$ rises further—signifying the presence of an optimal central density at which the configuration achieves hydrostatic equilibrium and maximum stability. Beyond this point, the configurations become dynamically unstable and are prone to collapse under their own gravity, in agreement with standard compact-star stability analyses based on the turning-point criterion~\cite{Tooper1964,Glendenning2000}. 
The monotonic decrease in radius with increasing $\epsilon_{0\rm WD}$ reflects the enhanced gravitational binding and compactness at higher densities. Furthermore, the rate of decline in radius becomes steeper for higher DM fractions ($f_{\rm DM} \geq 0.08$), indicating that DM not only reduces the maximum attainable mass but also shifts the stability boundary to lower central energy densities. 
These results are consistent with prior studies that demonstrated similar dependencies of mass and radius on central energy density in DM-admixed compact stars~\cite{leung13,carvalho25,parmar25}. The observed trends quantitatively capture the delicate balance between degeneracy pressure (which increases with central density) and the gravitational pull (enhanced by DM presence), ultimately governing the equilibrium and stability of WD–DM configurations. The combined analysis of $M \sim n_{T0}$ and $M \sim \epsilon_{0\rm WD}$ relations demonstrates that even a modest DM admixture ($f_{\rm DM} \approx 0.05$) or a moderately heavy DM particle ($m_{\rm DM} \gtrsim 5~\mathrm{GeV}$) can significantly alter the global structure of white dwarfs—reducing both their maximum mass and radius. These quantitative effects are consistent with the predictions of Refs.~\cite{leung13,carvalho25,parmar25} and suggest that precise observational measurements of WD radii and masses could serve as potential probes for constraining the nature and fraction of dark matter captured in compact stellar objects.
\begin{figure*}[t]
\centering
\includegraphics[width=0.95\textwidth]{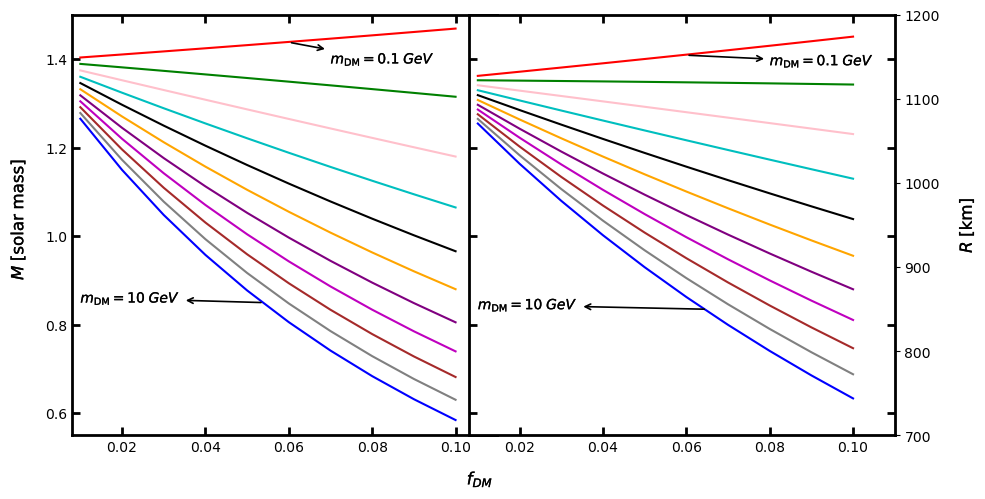}
\caption{Dependence of WD-DM mass and radius on the dark matter mass fraction $f_\mathrm{DM}$ for different dark matter particle masses. Left panel: The variation of stellar mass with $f_\mathrm{DM}$. Right panel: The corresponding variation of stellar radius with $f_\mathrm{DM}$. The arrows indicate the curves corresponding to $m_\mathrm{{DM}}=\mathrm{0.1\;GeV}$ and $m_\mathrm{{DM}}=\mathrm{10\;GeV}$, while the intermediate curves represent other dark matter particle masses not explicitly marked.}
\label{fig7}
\end{figure*} 
\begin{figure*}[t]
\centering
\includegraphics[width=1.01\textwidth]{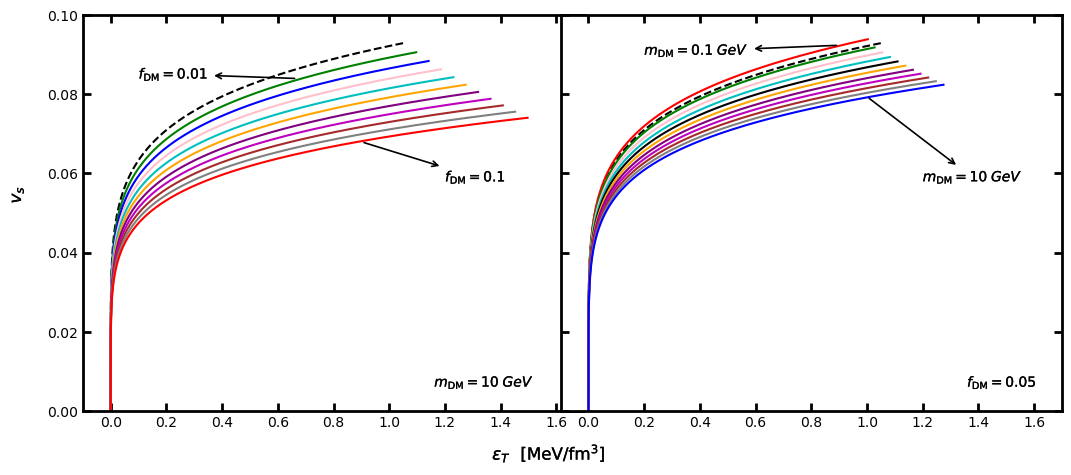}
\caption{Variation of the speed of sound ($v_{s}$) with total energy density ($\epsilon_{T}$) in dark matter-admixed white dwarfs. Left panel: The behavior of $v_{s}$ for a fixed dark matter particle mass $m_\mathrm{{DM}}=\mathrm{10\;GeV}$ with varying dark matter fractions $f_\mathrm{DM}$, where the arrows indicate the curves corresponding to $f_\mathrm{{DM}}=0.01$ and $0.1$. Right panel: The variation of $v_{s}$ for a fixed dark matter fraction $f_\mathrm{{DM}}=0.05$ with varying $m_\mathrm{DM}$, where arrows mark $m_\mathrm{{DM}}=\mathrm{0.1\;GeV}$ and $\mathrm{10\;GeV}$. Intermediate curves correspond to unlabeled intermediate values. The dashed black line denotes the standard Chandrasekhar white dwarf without dark matter. }
\label{fig8}
\end{figure*} 
\begin{figure*}[t]
    \centering
    \includegraphics[width=0.95\textwidth]{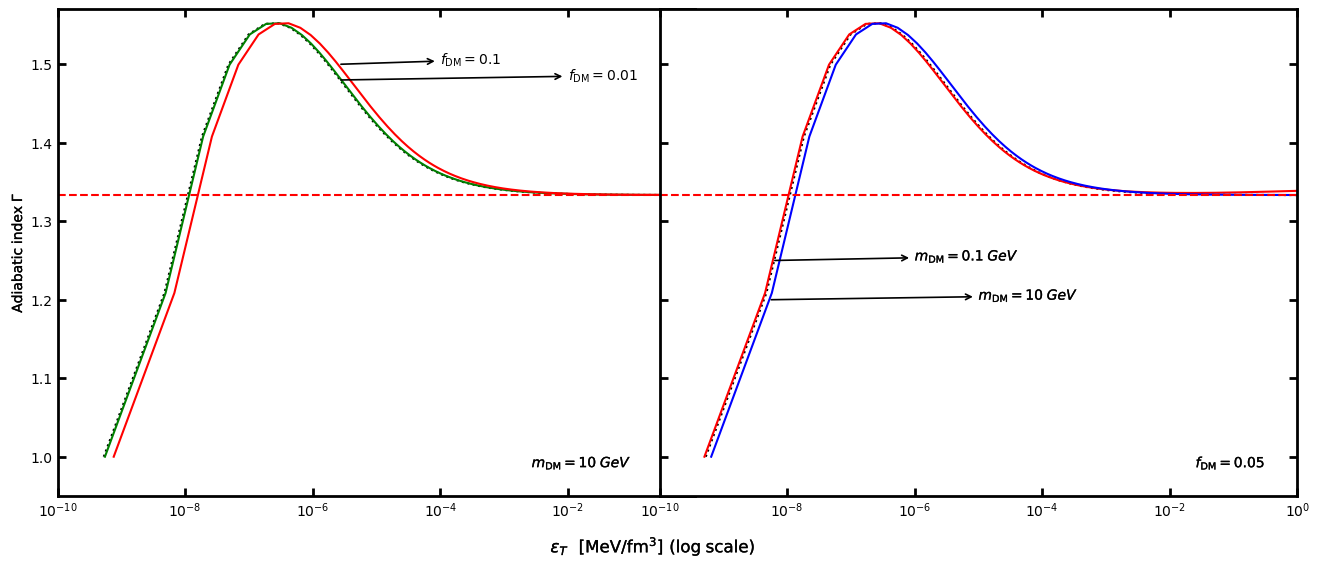}
    \caption{Variation of the adiabatic index ($\Gamma$) as a function of total energy density ($\epsilon_{T}$, in log scale) for dark matter-admixed white dwarfs. Left panel: $\Gamma$ for fixed dark matter particle mass $m_\mathrm{{DM}}=\mathrm{10\;GeV}$ with varying dark matter fractions $f_\mathrm{DM}$; arrows indicate $f_\mathrm{DM}=0.01$ and $0.1$, while intermediate curves are omitted for clarity. Right panel: $\Gamma$ for fixed dark matter fraction $f_\mathrm{{DM}}=0.05$ with varying $m_\mathrm{DM}$; arrows mark $m_\mathrm{{DM}}=\mathrm{0.1\;GeV}$ and $\mathrm{10\;GeV}$, with intermediate curves not shown. In both panels, the dotted black line denotes the standard Chandrasekhar white dwarf without dark matter, and the red dashed line represents the critical stability threshold at $\Gamma = 4/3$. }
    \label{fig9}
\end{figure*} 
\subsection{Dependence of WD-DM Mass and Radius on the Dark Matter Fraction} 
The left and right panels of Fig.~\ref{fig7} present the variation of the total stellar mass and radius of dark matter-admixed white dwarfs (WD–DM systems) as functions of the dark matter fraction ($f_{\rm DM}$) for several values of the dark matter particle mass ($m_{\rm DM}$). The results clearly indicate that both the stellar mass and radius systematically decrease as $f_{\rm DM}$ increases. This behavior reflects the combined effects of equation of state (EoS) softening and enhanced gravitational binding due to the inclusion of a non-luminous, pressure-weak DM component. For a representative case with $m_{\rm DM}=1~\mathrm{GeV}$, the stellar mass decreases from $M \approx 1.4 \,M_\odot$ at $f_{\rm DM}=0$ (the Chandrasekhar limit) to $M \approx 1.36\,M_\odot$ at $f_{\rm DM}=0.05$, and further to $M \approx 1.32\, M_\odot$ at $f_{\rm DM}=0.1$. Correspondingly, the stellar radius contracts from $R \approx 1020 \, \mathrm{km}$ (for $f_{\rm DM}=0$) to $R \approx  1018\,\mathrm{km}$ and $R \approx 974\,\mathrm{km}$ for $f_{\rm DM}=0.05$ and $0.1$, respectively. These trends correspond to a $\sim 3-6\%$ reduction in mass and a $\sim 0.2-5\%$ reduction in radius, consistent with the expected EoS softening. As the total pressure $(P_\mathrm{T}=P_\mathrm{WD}+P_\mathrm{ DM}$) becomes increasingly dominated by the less degenerate DM component, the overall pressure support against gravitational collapse diminishes. A Comparison across different DM particle masses reveals that heavier DM particles produce significantly more compact configurations. At $f_{\rm DM}=0.05$, the total mass decreases from $M \approx 1.43\,M_\odot$ for $m_{\rm DM}=0.1~\mathrm{GeV}$ to $M \approx 1.36\,M_\odot$ for $m_{\rm DM}=1~\mathrm{GeV}$ and further down to $M \approx 0.88\,M_\odot$ for $m_{\rm DM}=10~\mathrm{GeV}$, while the corresponding radii change from $R \approx  935\,\mathrm{km}$ to $1018\,\mathrm{km}$ and $757\,\mathrm{km}$, respectively. This strong dependence on $m_\mathrm{DM}$ arises because the degeneracy pressure of fermionic DM scales as $P_\mathrm{DM} \propto n_\mathrm{DM}^{5/3}/m_{\rm DM}$, hence, larger particle masses $m_{\rm DM}$ values yield significantly weaker pressure for a given number density, resulting in more centrally condensed and gravitationally compact systems. These quantitative behaviors are in strong agreement with the findings of earlier investigations on dark matter admixed compact stars~\cite{leung13,carvalho25}. Hence the present analysis confirms both the DM fraction $f_{\rm DM}$ and the DM particle mass $m_{\rm DM}$ act as crucial control parameters governing the global structure of mixed stars across a wide range of compact objects from white dwarfs to neutron stars. Physically, this dependence indicates that increasing the DM mass fraction enhances the gravitational potential while simultaneously reducing the degeneracy pressure contribution per unit mass, leading to the formation of smaller and denser stars. 
\subsection{Stability Analysis of Dark Matter-Admixed White Dwarfs} 
To investigate the dynamical stability of the white dwarf–dark matter (WD–DM) configurations, we analyze the variation of the speed of sound ($v_s$) and the adiabatic index ($\Gamma$) within the stellar interior, as shown in Figs.~\ref{fig8} and~\ref{fig9}. These quantities provide direct insight into the stiffness of the total equation of state (EoS) and the system’s ability to resist radial perturbations. As depicted in Fig.~\ref{fig8}, $v_s$ increases monotonically with total energy density ($\epsilon_{T}$), reaching typical values of $v_s \approx 0.00\text{–}0.03$ in the degenerate outer layers and up to $v_s \approx 0.065\text{–}0.09$ near the stellar center, depending on the dark matter fraction and particle mass. These values lie well below the causal limit ($v_s < 1$), ensuring that the inclusion of a DM component does not lead to any violation of causality or introduce thermodynamic instability. 
The increase of $v_s$ with $\epsilon_{T}$ reflects the gradual stiffening of the matter under strong compression, consistent with the behavior of standard degenerate matter.
The adiabatic index, defined as $\Gamma = \frac{\epsilon_{T} + P_{T}}{P_{T}} \left( \frac{dP_{T}}{d\epsilon_{T}} \right)$, characterizes the local stability of the configuration against adiabatic perturbations. A stellar configuration remains dynamically stable if $\Gamma > 4/3$ throughout most of the stellar interior~\cite{Chandrasekhar1964}. As shown in Fig.~\ref{fig9}, the value of $\Gamma$ remains above the critical threshold ($\Gamma \geq 1.33$) across the majority of the stellar volume for all DM mass fractions and particle masses considered in this work. For example, at the stellar center ($r \approx 0$), $\Gamma$ typically lies in the range $1.45 \lesssim \Gamma \lesssim 1.55$ for the case $f_{\rm DM} = 0.05$ and $m_{\rm DM}=10~\mathrm{GeV}$. A mild drop in $\Gamma$ occurs near the stellar surface due to the rapid decline in pressure, but this region contributes negligibly to the total gravitational binding energy and thus does not affect global stability of the configuration. Overall, our results confirm that a moderate dark matter admixture--characterized by $f_{\rm DM} \lesssim 0.1$ and $m_{\rm DM} \lesssim10~\mathrm{GeV}$, preserves both causality and hydrodynamic stability. Thus, the WD-DM configurations explored in this work represent physically viable stellar models within the single-fluid framework. 
\section{Summary and Conclusions} \label{sec:summary}
In this work, we investigated the equilibrium and stability properties of white dwarfs admixed with fermionic dark matter (DM) using, for the first time, a single-fluid formalism. The total equation of state (EoS) was constructed by combining the baryonic and DM contributions for a wide range of DM fractions $f_{\rm DM}$ and DM particle masses $m_{\rm DM}$. The Higgs-portal interaction modifies the internal thermodynamics by reduced the effective pressure at fixed energy density. Quantitatively, increasing either $f_{\rm DM}$ and/or $m_{\rm DM}$ softens the EoS—typically lowering the pressure by $2-9\%$ and producing more compact stellar configurations. Solutions of the TOV equations reveals that both the maximum mass and radius decrease monotonically with either DM parameter. For fixed $m_{\rm DM}$, raising the DM fraction from $f_{\rm DM}=0$ to $0.1$ reduces the maximum mass by $\sim 3-34\%$ and the radius by $\sim 2-23\%$. For fixed $f_{\rm DM}$, increasing the DM particle mass from $1\, \rm{to}\, 10$ GeV produces an even stronger suppression, with maximum masses reduced by up to $\sim 20\%$. The mass–central density relation exhibits the standard relativistic behavior: configurations with $dM/d\epsilon_{T0}>0$ are stable, while the turning point marks the onset of gravitational instability. Radii decrease monotonically with increasing central energy density, and the inclusion of DM accelerates this compactification trend.
Stability tests further confirm the physical viability of the configurations. The sound speed satisfies causality ($v_s<1$) for all parameter sets considered. The adiabatic index remains above the critical value ($\Gamma>4/3$) throughout the mass-carrying interior, dropping below $4/3$ only in negligible low-density surface layers. Thus, dynamical stability is preserved along the stable branch. However, when $f_{\rm DM}\gtrsim 0.1$ and $m_{\rm DM}\gtrsim 10\,\mathrm{GeV}$, the maximum-mass point shifts to significantly lower central densities, indicating that the WD–DM system becomes gravitationally unstable and prone to collapse. Our results demonstrate that even a modest DM admixture can significantly alter observable properties of white dwarfs. Future work should extend the analysis to a two-fluid formalism, incorporate rotation and magnetic fields, and explore self-interacting or non-thermal DM scenarios, which may have important implications for WD cooling, oscillations, and potential gravitational-wave signatures.
\section*{Acknowledgements}
Mrutunjaya Bhuyan acknowledges support from the Anusandhan National Research Foundation through the Ramanujan Fellowship, File No. RJF/2022/000140.

\bibliographystyle{JHEP}      
\bibliography{bibtex}
\end{document}